\documentstyle[12pt]{article}
\title{The role of acceleration and locality in the twin paradox}
\author{Hrvoje Nikoli\'c  \\
Theoretical Physics Division, Rudjer Bo\v{s}kovi\'{c} Institute, \\
P.O.B. 180, HR-10002  Zagreb, Croatia \\
{\normalsize hrvoje@faust.irb.hr} \\
\makebox[1in]{} \\
}
\date{\today}
\begin{document}
\maketitle
\begin{abstract}
We study the role of acceleration in the twin paradox. From the coordinate 
transformation that relates an accelerated and an inertial 
observer we find that, from the point of view of the accelerated observer, 
the rate of the differential lapses of time
depends not only on the relative velocity, but also on the
product of the acceleration and the distance between the observers. 
However, this result does not have a direct operational interpretation
because an observer at a certain position can measure only physical 
quantities that are defined at the same position. 
For local measurements, the asymmetry between the two 
observers can be attributed to the
fact that noninertial coordinate systems, contrary to inertial coordinate
systems, can be
correctly interpreted only locally.  
\end{abstract}

\section{Introduction}

According to the special theory of relativity, all motions with a constant 
velocity are relative. The twin paradox consists in the fact that 
a twin $B$ that travels around and eventually meets his brother $A$ is younger 
than his brother $A$. How the twin $B$ knows that he is the one who is actually 
moving? An often answer, especially in the older literature, 
is: ``He knows, because he accelerates and  
consequently feels an inertial force." However, it has been stressed many times 
in the literature$^{(1-5)}$ that acceleration is not an essential part of the twin 
paradox. 

The most general explanation of the twin paradox is purely geometrical; 
the proper lengths of two trajectories in spacetime, which correspond  
to times measured by the corresponding observers, do not need to be 
equal. This explanation is perfectly correct and is also the simplest one, 
because one does not need an explicit coordinate transformation that 
relates the two observers. Yet, such an explanation may not be 
completely satisfactory, because one may want to know  
where and when the different aging of the two observers occurs. Although, 
strictly speaking, this question is not really meaningful,$^{(4)}$ we show 
in Sec. 2 that, in a certain sense, the different aging can be attributed 
to instants of time when one of the observers accelerates. We find 
that the accelerated observer ``observes" that the rate of the 
differential lapses of time  
depends not only on the relative velocity, but also on the 
product of the acceleration and the distance between the observers. 
This result has already been obtained in Ref. 6, but our derivation is 
quite different, and, we believe, more elegant. However, we also 
emphasize that this result does not have a direct operational interpretation 
because an observer at a certain position can only measure physical 
quantities that are defined at the same position.      

Another way of posing the twin paradox is to ask what, if not 
acceleration, is the source of asymmetry between the two 
relatively moving observers. When spacetime is curved$^{(1)}$ or has a 
nontrivial topology,$^{(3,7,8)}$ one can obtain the twin paradox 
completely without acceleration. In these two cases it is clear 
that, owing to a nontrivial geometry or topology, there is no symmetry 
with respect to rotations of the velocity directions.
However, a flat opened universe possesses such a symmetry, so the 
question of the source of asymmetry for such a case remains 
opened. We find in Sec. 3 that this asymmetry can be attributed to the 
fact that noninertial coordinate systems, contrary to inertial coordinate
systems, can be 
correctly interpreted only locally. Recently, this fact has been used 
to resolve the Ehrenfest paradox,$^{(9)}$ to give the correct 
interpretation of the Sagnac effect,$^{(9)}$ and to show that 
the notion of radiation does not depend on acceleration of 
an observer.$^{(10)}$ In this paper we explain how this fact helps in 
understanding the twin paradox.    

\section{The role of acceleration}

Let $S$ be the frame of an inertial observer and $S'$ the frame of an observer 
that moves arbitrarily along the $x$-axis. $S$ is a Lorentz frame, while 
$S'$ is actually a coordinate system determined by the Fermi-Walker transport along 
the trajectory of the arbitrarily moving observer. (We give more comments on 
this in Sec. 3). 

Let $u$ be the velocity of the observer 
in $S'$ as seen by an observer in $S$. Let us also assume that the observers 
in $S$ and $S'$ do not rotate (this makes the analysis simpler, but has no 
influence on the final results). The coordinate transformation between 
these two frames is given by$^{(11,9)}$ $y=y'$, $z=z'$, and 
\begin{equation}\label{1}
x=\gamma(t')x' +\int_{0}^{t'}\gamma(t')u(t') dt' \; ,
\end{equation}
\begin{equation}\label{2}
t=\frac{1}{c^2}\gamma(t')u(t')x' +\int_{0}^{t'}\gamma(t')dt' \; ,
\end{equation}
where $\gamma(t')=1/\sqrt{1-u^2 (t')/c^2}$. 
It was assumed in this transformation that  
the space origins coincide at $t=t'=0$. The 
position of the observer in $S'$ is $x'=0$. 

%In Ref. 14 it is given a recipe how to find the explicit form of 
%$u(t')$ for a general case, where the twin paradox for a uniform 
%acceleration has been investigated as a byproduct. However, the 
%roles of position, velocity and acceleration are more manifest in 
%a general analysis than in a special example, so we proceed with 
%the general analysis.    

The transformation (\ref{1})-(\ref{2}) 
is linear in $x'$. However, if $u(t')$ is not a constant, then 
this transformation is not linear in
$t'$. Contrary to the case of constant $u$, this transformation 
cannot be simply inverted 
by putting $u\rightarrow -u$. This is why the inertial and 
the noninertial 
observers are not equivalent. Note, however, that acceleration does not appear
explicitly in (\ref{1}) and (\ref{2}).  

Let us now see how the clock in $S'$ appears to the observer in $S$. Since 
the clock in $S'$ is at $x'=0$, from (\ref{2}) we find
\begin{equation}\label{i1}
t=\int_{0}^{t'}\gamma(t')dt' \; ,
\end{equation}
\begin{equation}\label{i2}
\frac{\partial t}{\partial t'}=\gamma(t') \; .
\end{equation}
Equations (\ref{i1}) and (\ref{i2}) express the fact that the observer in $S$
sees that the clock in $S'$ is slower than the clock in $S$ and that 
$\partial t/\partial t'$ does not depend on the acceleration, but only on 
the instantaneous velocity. 
  
Let us now see how the clock in $S$, not necessarily at $x=0$, 
appears to the observer in $S'$. 
By eliminating $x'$ from (\ref{1}) and (\ref{2}), we find 
\begin{equation}\label{3}
t=\int_{0}^{t'}\gamma(t')dt' +\frac{1}{c^2}u(t')[x-x_o(t')] \; ,
\end{equation}
where
\begin{equation}\label{3.1}
x_o(t')=\int_{0}^{t'} \gamma(t')u(t')dt'=\int_{0}^{t(t')} u(t'(t))dt 
\end{equation}
is the position of the observer in $S'$ as a function of time, 
$t(t')$ is given by (\ref{i1}), and  
$t'(t)$ is its inverse. It was allowed to use (\ref{i1}) and (\ref{i2}) 
in the second equality in (\ref{3.1}) because, although $u$ is expressed 
as a function of $t'$, it is, by definition, the velocity seen by the 
$S$-observer, i.e., $u=dx_o/dt$. 
From (\ref{3}) and (\ref{3.1}) 
we find
\begin{equation}\label{4}
\frac{\partial t}{\partial t'}=\frac{1}{\gamma(t')} +\frac{1}{c^2}
\frac{du(t')}{dt'}[x-x_o(t')] \; . 
\end{equation}
The fact that (\ref{4}) and (\ref{i2}) are different is a consequence 
of the fact that 
the partial derivative in (\ref{4}) is calculated with $x$ held fixed, 
while the partial derivative in (\ref{i2}) is calculated with $x'$ held
fixed.   
Comparing (\ref{4}) with (\ref{i2}), we see that the first term 
in (\ref{4}) corresponds to what we expect from the relativity of motion. 
However, the second term in (\ref{4}) shows that acceleration has a direct 
influence on what the accelerated observer will observe. Note also that the 
influence of acceleration does not depend only on the acceleration itself, but 
also on the relative distance between the accelerated observer and the 
inertial clock. In particular, if the inertial and the noninertial clocks 
are at the same instantaneous position, then acceleration has no influence. 

Now it seems that we understand what is the true origin of the different 
aging of the inertial and the noninertial clocks. For example, if the observer in 
$S'$ moves with a constant velocity and then suddenly reverses the direction 
of motion, then, at this critical instant of time, it will appear to him 
that the time of the inertial clock at $x=0$ instantaneously jumps forward. 
There is no such jump of the noninertial clock 
from the point of view of the inertial observer.  
This is the reason that, when the two observers finally meet, they have 
different age. A similar conclusion, although obtained in a 
completely different 
way, was drawn also in Ref. 13. However, as we show in the next section, 
this is not the end of the story.  

\section{The role of locality}

The discussion of the preceding section seems to resolve the twin 
paradox. However, this discussion raises a new paradox.  
The right-hand side of (\ref{4}) can be negative, which means that it may 
appear to the accelerated observer that an inertial clock lapses 
backward in time. This seems to be in contradiction with the 
principle of causality. 

This paradox is an artefact of the tacit assumption that an observer 
receives information from a distant clock {\em instantaneously}. 
In that sense, equations (\ref{i1}), (\ref{i2}), (\ref{3}), and (\ref{4}) 
do not represent what the observers will really see, unless the clocks 
and the observers 
in $S$ and $S'$ are at the same instantaneous position. One could 
caculate what the observers would really see by assuming that the observers 
communicate with light signals, which would remove the causality paradox. 
However, we do not want to introduce a new entity, such as a light beam 
needed for communication, because such a complication could hide the 
real origin of the twin paradox. Instead, we insist on resolving the 
twin paradox using only the properties of the transformation  
(\ref{1})-(\ref{2}). 

An observer at a certain position can measure only the values of physical 
quantities at this same position. Therefore, equations (\ref{3}) and
(\ref{4}) have a direct operational interpretation only for $x=x_o(t')$. 
Therefore, when the observer in $S'$ compares his clock with a clock in 
$S$ at the same instantaneous position, he sees
\begin{equation}\label{5}
t=\int_{0}^{t'}\gamma(t')dt' \; ,
\end{equation}
\begin{equation}\label{6}
\frac{\partial t}{\partial t'}=\frac{1}{\gamma(t')} \; .
\end{equation}
The apparent inconsistency of equations (\ref{5}) and (\ref{6}) is a new 
way of viewing the twin paradox. Equations (\ref{6}) and (\ref{i2})  
correspond to the relativity of motion; the two observers do not agree 
on which clock is faster.  
On the other hand, equations (\ref{5}) and (\ref{i1}) correspond to 
the twin paradox; the two observers {\em do} agree that, at the same 
instant and the same position, the 
hand of the clock in $S'$ points to a smaller number than the hand of the clock
in $S$. But how is that possible? 

Note first that the apparent inconsistency of 
(\ref{5}) and (\ref{6}) has a simple mathematical origin. One should 
not derive (\ref{6}) directly from (\ref{5}), but instead one needs 
{\em first} to calculate the derivative of (\ref{3}) and {\em then} 
to put $x=x_o(t')$. However, (\ref{5}) and (\ref{6}) are correct 
equations even for a motion with a constant velocity. What is the 
source of asymmetry between $S$ and $S'$?      

The asymmetry lies in another tacit assumption; the condition 
$x=x_o(t')$ corresponds to an experimental arrangement in which 
the moving clock in $S'$ is compared each time with {\em another}
clock in $S$, that which, at this instant, is at the same position 
as the clock in $S'$. In other words, there is precisely one clock 
in $S'$, while there are many clocks in $S$ distributed along 
the $x$-axis. 

One will say: ``OK, but we can arrange our experiment such that 
there is only one clock in $S$, say at $x=0$, while there are 
many clocks in $S'$ distributed along the $x'$-axis." However, now comes 
the essential point of this section. It was legitimate to use 
many clocks at different positions in $S$ and compare the 
clock in $S'$ each time with another clock in $S$, because all these 
clocks in $S$ belong to the {\em same} frame of reference. 
On the other hand, if, at least for a short time,  
$S'$ is not an inertial frame,  
then we cannot longer say that clocks 
at different constant positions $x'$ belong to the same
frame of reference. In other words, the proper coordinates 
related to the trajectory $x'={\rm constant}$ depend on this constant. 
Consequently, it is not legitimate to compare 
the inertial clock at $x=0$ each time with another noninertial 
clock that does not move with respect to the noninertial 
clock at $x'=0$ and interpret the result as something that tell us 
about the behavior of the clock at $x'=0$. In the context of the twin 
paradox, a privileged role of 
inertial coordinate systems is not related to the fact that inertial observers 
do not feel an inertial force, but rather to the fact that inertial
coordinate systems  
in flat spacetime have a clear {\em global} interpretation, while 
noninertial coordinate systems only have a clear {\em local} interpretation. 
This purely local interpretation of noninertial coordinate systems 
has already been explained in more detail  
in Refs. 9. and 10., but, for the sake of completeness, below we give a short 
resume of the results of these papers.              

Proper coordinates of an observer arbitrarily moving in arbitrary 
spacetime are determined by the Fermi-Walker transport $^{(12)}$ 
and are often referred to as Fermi coordinates.$^{(15)}$ 
They are determined 
by the trajectory of the observer, as well as by the geometry of 
spacetime. It is convenient to define them such that the position 
of the observer is at the space origin of the Fermi coordinates.  
Even if there is no relative motion between two observers, 
they belong to different Fermi frames if they are not staying at the same 
position. In particular, the $S$-coordinates in (\ref{1})-(\ref{2}) 
are the Fermi coordinates of an inertial observer in flat 
spacetime, while the $S'$-coordinates are the Fermi coordinates of an 
observer that moves arbitrarily (without rotation) along the $x$-axis 
with respect to the inertial observer. In general, the coordinate 
transformation that relates the Fermi coordinates of two different 
observers is a complicated transformation. However, the coordinate 
transformation that relates the Fermi coordinates of two inertial observers 
in flat spacetime that do not move relatively to each other is a 
simple translation of the space origin, which is a transformation 
that belongs to the class of restricted internal transformations,$^{(9)}$ 
i.e., it is a transformation that does not mix space and time coordinates:  
\begin{equation}\label{rit}
t'=f^{0}(t) \; , \;\;\;\; x'^i =f^{i}(x^1,x^2,x^3) \; .
\end{equation}
If the Fermi coordinates of two observers are related by such a transformation, 
then they can be regarded as belonging to the same {\em physical} frame of 
reference. This is why one can regard the clocks at different positions 
$x$ in (\ref{3}) as belonging to the same frame of reference.    

Now we also have a better understanding of the causality ``paradox"
mentioned at the begining of this section; it is merely a consequence 
of the fact that the coordinate $t'$ can be interpreted as a physical 
time only at $x'=0$. 

At the end, 
let us also mention some measurable effects related to the local interpretation 
of noninertial frames. Assume that the $S'$-coordinates in (\ref{1})-(\ref{2}) 
refer to a uniformly accelerated observer at $x'=0$. Assume also that another 
observer has such a trajectory that his position is given by 
$x'=$constant$\neq 0$. Then this second observer also accelerates 
uniformly, but with a {\em different} acceleration.$^{(14)}$ If, on the other
hand, two observers at different positions move in the same direction 
with equal accelerations and initial velocities, then, 
as seen by these observers, the distance 
between them changes with time, which also leads to a variant of the 
twin paradox.$^{(2)}$ 

\section*{Acknowledgments}

I am grateful to G. Duplan\v{c}i\'{c} for asking me questions that 
motivated the investigation resulting in this paper.  
This work was supported by the Ministry
of Science and Technology of the
Republic of Croatia under Contract No. 00980102.

\section*{References}

%\mbox{ }\newline
 1. B. R. Holstein and A. R. Swift, %``The Relativity Twins in Free Fall,"  
 {\it Am. J. Phys.} {\bf 40}, 746 (1972).
\newline
 2. S. P. Boughn, %``The case of the identically accelerated twins," 
 {\it Am. J. Phys.} {\bf 57}, 791 (1989).
\newline
 3. T. Dray, %``The Twin Paradox Revisited," 
 {\it Am. J. Phys.} {\bf 58}, 822 (1990).
\newline
 4. T. A. Debs and M. L. G. Redhead, %``The twin ``paradox" and the
 %conven-
 %\newline\hspace*{0.2cm}tionality of simultaneity," 
 {\it Am. J. Phys.} {\bf 64}, 384 (1996).
\newline
 5. R. P. Gruber and R. H. Price, %``Zero time dilation in an accelerating 
 %rocket," 
 {\it Am. J. Phys.} {\bf 65}, 979 (1997).  
\newline
 6. R. J. Low, %``When Moving Clocks Run Fast," 
 {\it Eur. J. Phys.} {\bf 16}, 228 (1995).
\newline
 7. R. J. Low, {\it Eur. J. Phys.} {\bf 11}, 25 (1990).
\newline
 8. J.-P. Uzan, J.-P. Luminet, R. Lehoucq, and P. Peter, physics/0006039.
\newline
 9. H. Nikoli\'{c}, %``Relativistic contraction and related effects in 
 %noninertial frames," 
 {\it Phys. Rev. A} {\bf 61}, 032109 (2000). 
\newline
 10. H. Nikoli\'{c}, %``Notes on covariant quantities in noninertial 
 %frames and invari-
 %\newline\hspace*{0.2cm}ance of radiation in classical and quantum 
 %field theory," 
 gr-qc/9909035.
\newline
 11. R. A. Nelson, %``Generalized Lorentz transformation for an accelerated, 
 %\newline\hspace*{0.2cm}rotating frame of reference,"
 {\it J. Math. Phys.} {\bf 28}, 2379 (1987).
\newline
 12. C. W. Misner, K. S. Thorne, and J. A. Wheeler, {\it Gravitation} (W. H.
 \newline\hspace*{0.6cm} Freeman and Company, New York, 1995).
\newline
 13. W. G. Unruh, %``Parallax distance, time, and the twin ``paradox"," 
 {\it Am. J. Phys.} {\bf 49}, 589 (1981). 
\newline
 14. H. Nikoli\'{c}, %``Relativistic contraction of an accelerated rod," 
 {\it Am. J. Phys.} {\bf 67}, 1007 (1999).
\newline 
 15. J. L. Synge, {\it Relativity: The General Theory} (North-Holland Publishing
 \newline\hspace*{0.6cm} Co., New York, 1960).
\end{document}